\documentclass[%
 reprint,
superscriptaddress,
 amsmath,amssymb,
 aps,pre,
]{revtex4-2}

\usepackage{epsfig,dsfont,amssymb,amsmath,amsthm,amsfonts,amsbsy,mathrsfs}
\usepackage{graphicx}
\usepackage{color}
\usepackage{bm}
\usepackage{multirow}
\usepackage{natbib}
\usepackage{ulem}

\newcommand{\rr}{{\mathbf r}}

\newcommand{\BEQ}{\begin{equation}}
\newcommand{\EEQ}{\end{equation}}
\newcommand{\BEA}{\begin{eqnarray}}
\newcommand{\EEA}{\end{eqnarray}}

\newcommand{\qv}{\mathbf{q}}

\usepackage{tikz}
\usetikzlibrary{arrows,shapes,chains}

\begin{document}
\preprint{APS/123-QED}

\title{ Alignment interactions drive structural transitions in biological tissues}

\author{Matteo Paoluzzi}
\email{matteopaoluzzi@ub.edu}
\affiliation{Departament de Física de la Mat\`eria Condensada, Universitat de Barcelona, C. Martí Franqu\`es 1, 08028 Barcelona, Spain.}
\author{Luca Angelani}
\affiliation{ISC-CNR,  Institute  for  Complex  Systems, Piazzale  A.  Moro  2,  I-00185  Rome,  Italy.}

\affiliation{Dipartimento di Fisica, Sapienza Universit\`a di Roma
Piazzale  A.  Moro  2,  I-00185  Rome,  Italy.}

\author{Giorgio Gosti}
\affiliation{Center for Life Nano Science, Istituto Italiano di Tecnologia, Viale Regina Elena 291, I-00161, Rome, Italy.}
\author{M Cristina Marchetti}
\affiliation{Department of Physics, University of California Santa Barbara, Santa Barbara, CA 93106, USA}
\author{Ignacio Pagonabarraga}
\affiliation{CECAM Centre  Europ\'een  de Calcul Atomique et Mol\'eculaire, \'Ecole Polytechnique F\'ed\'erale de Lausanne (EPFL),
Batochime, Avenue Forel 2, 1015 Lausanne, Switzerland.}
\affiliation{UBICS University of Barcelona Institute of Complex Systems, Mart\'i i Franqu\`es 1, E08028 Barcelona, Spain.} 
\affiliation{Departament de Física de la Mat\`eria Condensada, Universitat de Barcelona, C. Martí Franqu\`es 1, 08028 Barcelona, Spain.}
\author{Giancarlo Ruocco}
\affiliation{Center for Life Nano Science, Istituto Italiano di Tecnologia, Viale Regina Elena 291, I-00161, Rome, Italy}
\affiliation{Dipartimento di Fisica, Sapienza Universit\`a di Roma
Piazzale  A.  Moro  2,  I-00185  Rome,  Italy.}

\date{\today}

\begin{abstract}
Experimental evidence shows that there is 
a feedback
between cell shape and cell motion. How this feedback impacts the collective behavior of dense cell monolayers remains an open question.
We investigate the effect of 
a feedback
that tends to align the cell crawling direction with cell elongation in a biological tissue model.
We find that the alignment interaction
promotes nematic patterns in the fluid phase that eventually undergo a non-equilibrium phase transition into a quasi-hexagonal solid. Meanwhile, highly asymmetric cells do not undergo the liquid-to-solid transition for any value of the alignment coupling. In this regime, the dynamics of cell centers and shape fluctuation show features typical of glassy systems. 
\end{abstract}

\maketitle

\section{Introduction \label{sec:Intro}}
Eukaryotic cells at high packing fraction organize themselves into confluent monolayers, develop collective motion, and trigger a variety of patterns that play a fundamental role in complex biological processes ranging from wound healing to metastasis invasion \cite{sunyer2016collective,Friedl17}.
Pattern formation in biological tissues involves length scales that are much bigger than the typical cell length. This observation 
 suggests that a coarse-grained model of biological tissues needs to take into account only a few key ingredients of the single-cell dynamics \cite{trepat2018mesoscale}. 
Different approaches have been developed during the last few decades to capture the large-scale behavior of biological tissues
\cite{trepat2018mesoscale,alert2020physical,camley2017physical,banerjee2019continuum}.

Experimental studies discovered that biological tissues show glassy dynamics, support viscoelastic response, and behave as a disordered soft material in the vicinity of jamming or glassy transition \cite{Angelini11,ManningPNAS,Garcia15,kasza2007cell,trepat2009physical,angelini2010cell,pawlizak2015testing,schotz2008quantitative,PhysRevX.10.011016,saraswathibhatla2021coordinated}. 
However, differently from particulate systems, 
cell shape anisotropy is the driver of the jamming transition in confluent monolayers \cite{Park15,Bi14,Bi15}.
Thus, cell shape and its fluctuations are important ingredients that have to be taken into account in a mesoscopic description.
Shape fluctuations can be introduced in different ways \cite{alert2020physical,camley2017physical,henkes2020dense}.
Among the other alternatives, Vertex and Voronoi models are successful coarse-grained descriptions
that have been tested against different experiments in the last few years
\cite{Park15,Bi2016,Bi15,GiavazziPaoluzzi2018,wang2020anisotropy,sharp2019inferring,sussman2018soft,Matthias,staple2010mechanics,epl,farhadifar2007influence,PhysRevE.103.022607}.
\begin{figure}[!t]
\centering\includegraphics[width=.35\textwidth]{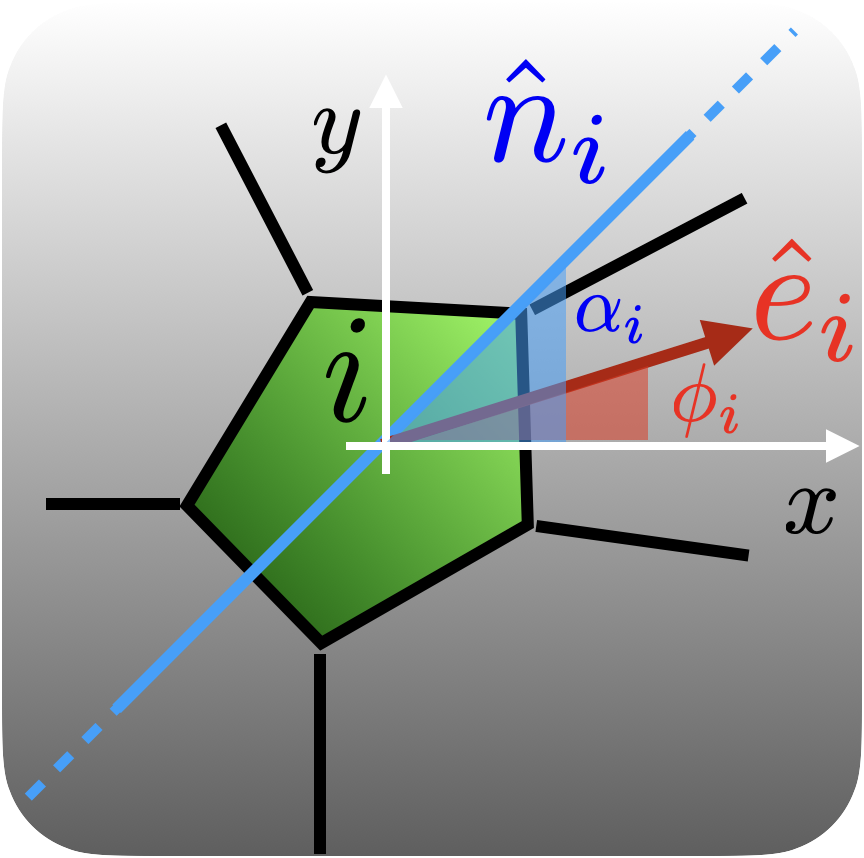}
\caption{ Pictorial representation of the model. The blue line represents the direction $\hat{\bf{n}}_i$ of the largest eigenvalue
of the shape tensor (the direction in the lab frame is parametrized by the angle $\alpha_i$). The red arrow is the self-propulsion direction $\hat{\bf{e}}_i$ (that is parametrized by the angle $\phi_i$). The alignment interaction tends to reduce the distance $\alpha_i - \phi_i$. 
}
\label{fig:model}      
\end{figure}

\begin{figure}[!t]
\centering\includegraphics[width=.5\textwidth]{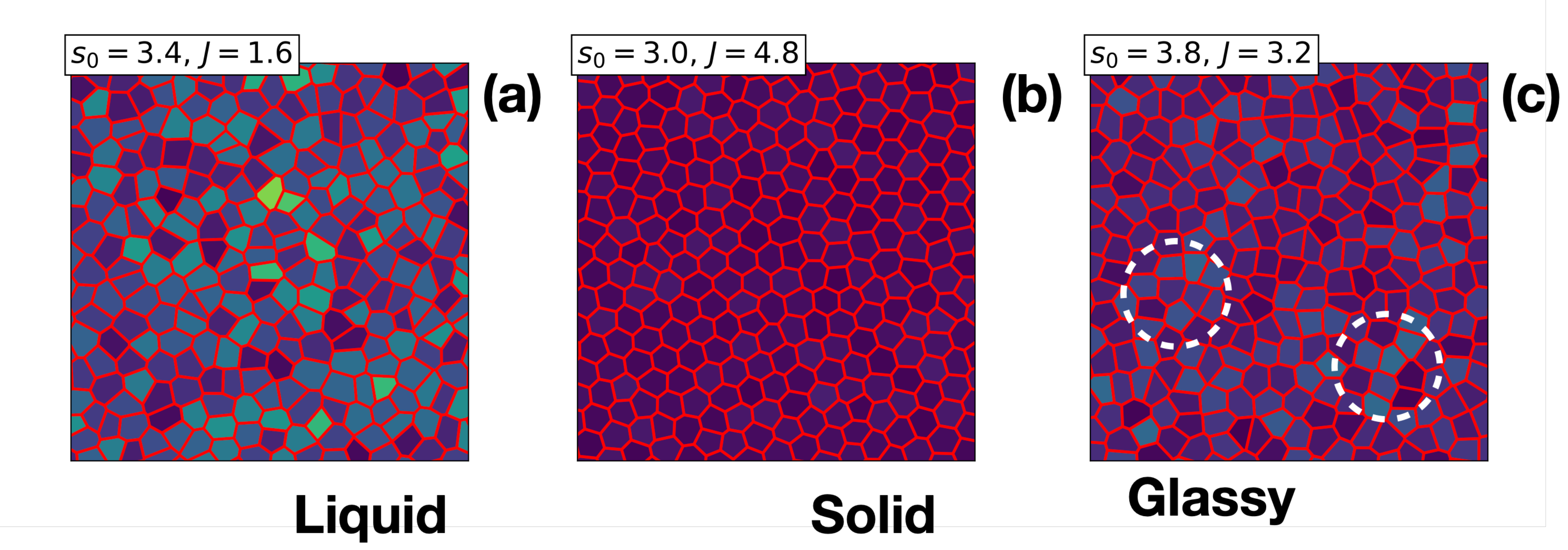}
\caption{ 
Representative snapshots of 
steady-state configurations taken in the weakly nematic liquid phase (a),
 disordered solid phase (b), and in the glassy regime (c).
The color indicates the modulus of the velocity in the lab frame from zero (dark) to its maximum value (yellow).
The alignment interaction acts as an inverse effective temperature:
As the strength of the interaction increases, velocity fluctuations become strongly inhibited promoting solidification. 
In the glassy regime, we observe the formation of dynamically heterogeneous regions (highlighted by the dashed white circles).
}
\label{fig:snap}      
\end{figure}

Because cells can move autonomously, biological tissue can be seen as a soft and active material \cite{Marchetti16}. 
It has been shown that feedback mechanisms at the single-cell level can trigger collective motion \cite{GiavazziPaoluzzi2018,Henkes11,petrolli2019confinement,czajkowski2018hydrodynamics}. 
Besides promoting collective migration, alignment interactions can also change the structural properties of a biological tissue \cite{GiavazziPaoluzzi2018}.
Structural changes and morphological transitions play a fundamental role in morphogenesis and organogenesis \cite{lancaster2014organogenesis}, however, the key ingredients responsible for self-organization and spatial differentiation in organoids are still poorly understood. 
Consequently, isolating the few fundamental ingredients that play the role of control parameters for the emergent structural organization
allows us to 
gain insight into complex
biological processes.

Cells become elongated during motion and tend to move along the direction of their long axis. As a consequence, cell motility is correlated with cell anisotropy \cite{maeda2008ordered,lauffenburger1996cell}.
Recent studies on phase-field models show that minimal dipolar interactions in monolayers of isotropic cells promote spontaneous symmetry breaking and nematic order \cite{mueller2019emergence}.

In this paper, we introduce a generalization of the Voronoi model
of biological tissues where we consider a minimal alignment interaction between cell shape and cell displacement.
We show that the
feedback between
shape and displacement 
triggers morphological transitions in 
the confluent monolayer.
The alignment interaction acts as an inverse effective temperature that
cools down the system as the intensity of the interaction increases.
Starting from fluid configurations,
the liquid becomes weakly nematic as the interaction is turned on. For higher values of the alignment interaction, the system falls into a hexagonal disordered solid \cite{hexatic1,hexatic2}.
We observe that the alignment interaction promotes the formation of cooperative clusters that tend to slow down the dynamics and trigger the proliferation of dynamical heterogeneities typical of glassy systems. Glassy dynamics involve both, the correlation of density fluctuations, as in the case of supercooled liquid, and shape fluctuations.

\section{Kinetic Monte Carlo Voronoi Model}
We implement a kinetic Monte Carlo (KMC) dynamics 
based on a Voronoi model of biological tissues \cite{Honda78}. 
The confluent monolayer is represented through the Voronoi tessellation of the $N$ cell centers (labeled by $i\!=\!1,...,N$, of coordinates $\mathbf{r}_i\!=\!(x_i,y_i)$, in a two dimensional square box of side $L\!=\!\sqrt{N}$ with periodic boundary conditions). 
 Let $\mathbf{r}\!\equiv\!(\mathbf{r}_1,..., \mathbf{r}_N )$ be a configuration of the system. 
The dynamics is governed by the following configurational energy \cite{Honda78,Nagai01,staple2010mechanics,Bi14,Bi15,Bi15a}
\BEQ \label{vertex}
E[\rr] = \sum_i \left[ K_a (a_i - a_0)^2 + K_p (p_i - p_0)^2 \right] \; ,
\EEQ
where the function $a_i $ and $p_i $ return the value of the area and the perimeter of the $i-$th polygon of the Voronoi tessellation. 
Cell area and cell perimeter fluctuate around the preferred (or target) values $a_0$ and $p_0$, their fluctuations are regulated by the stiffnesses $K_a$
and $K_p$. The square deviation from $a_0$
enforces the constraint of incompressibility in three
dimensions. The square deviation from $p_0$ encodes
the competition between cell-cell adhesion and active contractility in
the actomyosin cortex \cite{Bi15}. In the following we set $ a_0 \!=\! 1$, $K_p\!=\!K_a\!=\!1$, and we express (\ref{vertex})
in terms of the target shape index $s_0\!=\!p_0 / \sqrt{a_0}$ \cite{Bi15a}.
We sample stationary configurations of the energy
functional (\ref{vertex}) numerically 
using a Monte Carlo algorithm in which we propose 
time-correlated trial moves for
the cell centers. 
This algorithm is general enough to capture
two key features of the real cell movement dynamics:
(i) Cells usually move at a velocity that can fluctuate in magnitude \cite{vedel2013migration}, this
is ensured by the noise parameter $T$ that enters in the Monte Carlo algorithm, and
(ii) Cells displace positively correlated steps on a microscopic time scale $\tau$ \cite{dieterich2008anomalous,vedel2013migration}, 
which is the second parameter of the algorithm. 
Voronoi and Vertex models develop 
a rough energy landscape where energy 
barriers separate local minima \cite{Bi14,Bi15,Bi15a}. 
MC algorithms are particularly suitable 
for reaching steady-state configurations 
in these situations \cite{jorgensen1996monte,binder1995monte}.

Here the persistent motion typical of active systems is modeled using time-correlated trial displacements. We also consider an alignment interaction acting between the direction where the cell is elongated and the crawling direction.
Indicating with $\boldsymbol{\delta}_{i,t}$ the displacement
performed by the cell $i$ at the time step $t$, and adopting polar coordinates, we can write $\boldsymbol{\delta}_{i,t}=\delta_{i,t} (\cos \phi_{i,t},\cos \phi_{i,t})$, with $\delta_{i,t}=|\boldsymbol{\delta}_{i,t}|$. The angle $\phi_{i,t}$ determines the direction of the displacement. The shape tensor $\mathbf{Q}_i \!=\! (\sum_{l=1}^{N_v^i} \Delta \mathbf{r}_{l,i} \otimes \Delta \mathbf{r}_{l,i} ) / N_v^i $ encodes information on cell shape, where
$N_v^i$ is the number of cell $i$ vertices,
$\Delta \mathbf{r}_{l,i}=\mathbf{r}_{l,i} - \mathbf{r}_{CM,i}$ is cell $i$'s $l-$th vertex position,
$\mathbf{r}_{CM,i}$ indicates the center of mass,
and the symbol $\!\otimes\!$ indicates the standard diadic product.
In our case, $\mathbf{Q}_i$ is a $2 \times 2$ symmetric matrix.  
The eigenvector corresponding to the largest eigenvalue
defines the direction of maximum cell elongation.
We explore the effect of a nematic alignment interaction between the principal axis of elongated cell shape (parametrized through the angle $\alpha_i$) and the crawling direction (parametrized by the angle $\phi_i$). The alignment interaction is sketched in Fig. (\ref{fig:model}). 
For enforcing the alignment interaction,
at each MC step, we update the new direction $\phi_i$ using the following rule
\BEQ \label{eq:nematic}
\phi_{i,t+1} = \phi_{i,t} - J \sin 2 (\phi_{i,t} - \alpha_{i,t}) \; .
\EEQ
At the beginning of the time step $t\!+\!1$, the alignment interaction in Eq. (\ref{eq:nematic}) tends to align the (trial) displacement performed during the previous time step 
$\boldsymbol{\delta}_{i,t}$.  
Once updated the displacement direction, we propose the trial move 
\begin{align} \label{eq:deltar}
\mathbf{r}_{i,t+1}    \!=\! \mathbf{r}_{i,t} \!+\! \boldsymbol{\delta}_{i,t} \; , 
\end{align}
and then we evolve the displacement
\begin{align} \label{eq:deltadelta}
\boldsymbol{\delta}_{i,t} \!=\! \boldsymbol{\delta}_{i,t-1} \!+\! \delta_1 \boldsymbol{\eta}_{i}    
\end{align}
 with the condition $\boldsymbol{\delta}_{i,0} \!=\! \delta_0 \boldsymbol{\eta}_{i}$ \cite{LevisPRE}, i. e., in this way, the trial moves are correlated on a time scale $\tau \!=\! (\delta_0 / \delta_1)^2 \tau_{MC}$ \cite{BerhierMC,LevisPRE} 
(the Monte Carlo time step $\tau_{MC}$ is defined as the succession of $N$ elementary moves \cite{binder1995monte}). The components of the random vector $ \boldsymbol{\eta}_{i}$ are extracted  from a uniform distribution independently at each time step. The distribution is centered around zero and has unit variance.
Moreover, following Refs. \cite{Berthier11,LevisPRE}, the displacements are constrained to be $| \boldsymbol{\delta}_{i,t} | \!\leq\! \delta_0$ and $\delta_0 \!\geq\! \delta_1$ (in our simulations $\delta_0=0.25$).

The time evolution of the displacement $\boldsymbol{\delta}_{i,t}$ introduces a correlation on the time scale $\tau$ so that $\langle \delta_{i,t} \delta_{j,s}\rangle \sim \delta_{i,j}e^{-|t-s|/\tau}$ \cite{BerhierMC,LevisPRE}, as well as in the case of Active Ornstein-Uhlenbeck particles \cite{Maggi,Szamel15,Fily12,Fodor16,Paoluzzi}.
The model interpolates between an equilibrium relaxation dynamics for $\tau=0$, representing cells that perform a 
random crawling, to a persistent non-equilibrium dynamics characterized by a ballistic regime on short time scales, which is the hallmark of self-propelled motion at low Reynolds numbers. 
%
It is important to stress that, although the time evolution of $\phi_i$ given by Eq. (\ref{eq:nematic}) is deterministic, once we evolve $\boldsymbol{\delta}_{i,t}$ with Eq. (\ref{eq:deltadelta}), the stochastic term $\delta_1 \boldsymbol{\eta}_{i}$ introduces a rotational noise on $\phi_i$ that makes it to diffuse with a rotational diffusion constant $D_r\propto\tau^{-1}$. The trial move is thus accepted 
with probability $P_{acc} \!\propto\! \exp{(- \Delta E / T)}$,
with $\Delta E \equiv E[\mathbf{r}_{t+1}] - E[\mathbf{r}_{t}]$.

\begin{figure*}[!t]
\centering\includegraphics[width=1.\textwidth]{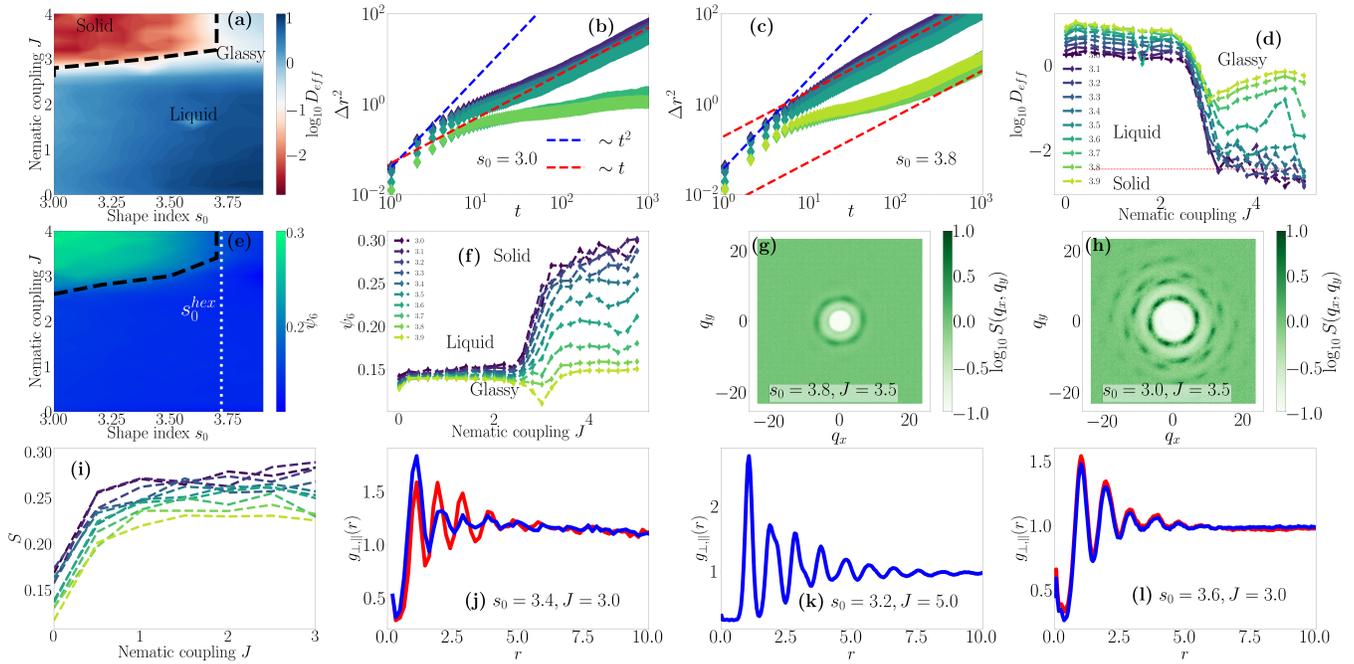}
\caption{ 
Structural properties. (a) phase diagram of the model 
using $D_{eff}$ as a dynamical order parameter. 
The dashed black line indicates the solid/liquid transition.
Crossing the liquid-solid transition 
the mean-squared displacement changes discontinuously, 
as shown in panel (b) 
($J\in[0,4]$ from violet to green, and $s_0=3.0$).
Approaching the glassy region, the mean-squared displacement shows a subdiffusive regime on intermediate times, as shown in (c)
for $s_0=3.8$ (same values of $J$ shown in (b)). 
The dashed red line is the diffusive scaling $\Delta r^2 \sim t$, dashed blue line the ballistic scaling $\Delta r^2 \sim t^2$.
(d) $D_{eff}$ as a function of the nematic coupling $J\in[0,4]$ for different values of $s_0 \in [3.0,3.9]$, increasing values from violet to yellow.
(e) Phase diagram using $\psi_6$ as a structural order parameter (increasing values of $\psi_6$ from blue to green).
The dashed black line represents the transition line to the hexagonal solid. The dotted white line corresponds to the value $s_0=s_0^{hex} \sim 3.722$.
(f) Order parameter $\psi_6$ as a function of $J$ for different values of $s_0\in[3.0,3.9]$, increasing values from violet to yellow.
Static structure factor $S(q_x,q_y)$ for $s_0=3.0$ (g) and $s_0=3.6$ (h). 
(i) Nematic order parameter $S$ 
as a function of $J$
(increasing values of $s_0\in[3.0,3.9]$ from violet to yellow). 
(j-l) Radial distribution function $g_{\parallel,\perp}(r)$ 
evaluated along $g_{\parallel}(r)=g(r_\parallel,0)$ and perpendicular  $g_{\perp}(r)=g(0,r_\perp)$
 the nematic director $\mathbf{n}$ ((j) liquid, (k) solid, and (l) glassy).}
\label{fig:pd}
\end{figure*}
We perform simulations of a tissue composed of $N\!=\!100,400,1600,6400$ cells with $T\!=\!0.002,0.05$ and $\tau\!=\!200$ (in Monte Carlo unit $\tau_{MC}$).

\section{Phase diagram}
We study the phase diagram of the tissue
using as control parameters $s_0$ and $J$.
The target shape index $s_0$ tunes the typical cell asphericity, i.e., the larger is $s_0$ the more elongated is the cell \cite{Bi15}.
We anticipate that the system shows a liquid-solid transition that is driven by the alignment coupling $J$. 
We perform numerical simulations in a region of the phase diagram where the system at finite temperature $T$ behaves as a fluid for any values of $\tau$ at $J\!=\!0$ (the phase diagram in the $T$ vs $\tau$ plane for $J\!=\!0$ is shown in Appendix (\ref{app:PDJ0})).

In Fig. (\ref{fig:snap}) we report three representative snapshots taken in the liquid, solid, and glassy regime. 
We observe an increase in velocity fluctuations as $s_0$ increases (see the Appendix (\ref{app:coll})). 
On the other hand, $J$ triggers
the formation of islands of slow-moving cells, as shown in Fig. (\ref{fig:snap}).
For $J\!\neq\!0$, 
the system develops nematic patterns as signaled by a non-vanishing value of the nematic order parameter $S$.
We monitored the average cell anisotropy via the parameter $\Delta = \langle \frac{(\lambda_1^i - \lambda_2^i)^2}{(\lambda_1^i+\lambda_2^i)^2}\rangle $, where the eigenvalues of the shape tensor $\lambda_{1,2}^i$ are sorted in a way so that $\lambda_1^i > \lambda_2^i$.
Our analysis reveals a jump to lower values of $\Delta$ crossing the liquid-to-solid transition. We anticipate that, in the solid phase, the system arranges into hexagonal patches with $\Delta \neq 0$ but small. Moreover, from the study of the relaxation dynamics of the principal axis, we obtain that axis fluctuations decorrelate on a finite time scale and no flipping dynamics between the two principal directions occurs (see the Appendix (\ref{app:CellAni})). These findings show that
the alignment interaction is always well defined.

\section{Structural Properties}
We start our quantitative discussion
from the phase diagram 
which is shown in Fig. (\ref{fig:pd})-(a). The phase diagram 
has been obtained considering the 
long time behavior of the 
mean-squared displacement $\Delta r^2$ (defined in the Appendix (\ref{app:obs})) as a dynamic order parameter \cite{Bi15}, i. e., looking at the effective diffusion constant $D_{eff} \!\equiv\! \lim_{t \to \infty} \Delta r^2 / (4t)$ \cite{Bi15,GiavazziPaoluzzi2018}. 
The typical behavior of $\Delta r^2$ is shown in (b) and in (c) 
 for $s_0\!=\!3.0,3.8$.  
As one can see, 
$\Delta r^2$ undergoes a crossover from
a ballistic regime on short time scales, \textit{i.e.}, $\Delta r^2\!\sim\!t^2$,  to a diffusive regime, \textit{i.e.}, $\Delta r^2 \!\sim\! 4 D_{eff} t$,
for longer times. As $J$ increases in intensity, we observe two different behaviors in $\Delta r^2$.
For small $s_0$ values (b), $\Delta r^2$ discontinuously develops a plateau right after the ballistic regime. 
This fact signals a liquid-solid transition. 
For larger $s_0$ values (c),
the plateau is replaced by a subdiffusive regime.
Panel (d) shows $D_{eff}$
as a function of $J$. 
Through this analysis, we identify three regimes in the phase diagram (see panel (a)):
a liquid phase for small $J$ values, a solid-state, at larger $J$ values and small $s_0$, and a glassy regime,  where $\Delta r^2$
develops a subdiffusive behavior at intermediate times. 

To gain insight into the structural properties of the system,
we take a look at the positional and orientational order. 
We start our discussion with the order parameter $\psi_6$ (see the Appendix (\ref{app:obs}) for its definition) for revealing the presence of sixfold order.
$\psi_6$ leads to the
phase diagram that is shown 
in Fig. (\ref{fig:pd}-e), where the color map indicates the magnitude of the order parameter.
The behavior of $\psi_6$ indicates that the solid phase is characterized by hexagonal order (see Fig. (\ref{fig:pd}-f))). 
In agreement with early studies on Vertex models \cite{staple2010mechanics,farhadifar2007influence}, 
for higher $J$ values, the transition between a glassy fluid and a disordered hexatic solid phase matches the critical value of a regular hexagon $s_0 \!=\! s_0^{hex}\!\sim\! 3.722$. 
Complementary information about the positional order is provided by the static structure factor $S(q_x,q_y)$ which allows us allows to visualize the emerging ordered 
patches. In Fig. (\ref{fig:pd})-(g,h) we report 
$S(q_x,q_y)$ in the glassy (g) and in the solid (h) phase. 
The solid phase shows hexatic patches that are compatible with a disordered hexagonal solid. The increase in positional order in the solid regime
is signaled by marked damped oscillations in the $g(r)$ that imply the lack of a true crystalline structure ($g(r)$ is reported in the Appendix (\ref{app:static})).
The order parameter $\psi_6$ jumps almost discontinuously at the transition (see panel (f)) providing evidence for an increase of hexatic order in the solid phase rather than in the liquid \cite{hexatic1,hexatic2}.


The control parameter that triggers the transition between liquid and solid is the alignment coupling $J$ 
which plays the role of an (inverse) effective temperature.
For rationalizing this effect,
we consider the simplest case where each
cell is represented by a
self-propelled spheroid undergoing an active Brownian dynamics with
self-propulsion velocity $v_0$, and rotational diffusion $\tau^{-1}$. 
During the dynamics, each particle tends to (i) minimize the mechanical energy, and (ii) align towards the direction given by $\alpha_i$.
We indicate with $\mathbf{r}^0=(\rr_1,\ldots,\rr_N)$ the inherent state configuration that minimizes $E[\mathbf{r}]$, \textit{i.e.}, $\left. \nabla E \right|_{\rr = \rr^0}=0$, and
we linearize the dynamics around those minima \cite{Henkes2018,Bi2016}. The equations of motion for the fluctuations are $\dot{\delta \mathbf{r}_i} \!=\! v_0 \mathbf{e}_i - \mu \mathbf{M}_{ij} \delta \mathbf{r}_j$, with $\delta \mathbf{r}_i\!=\!\mathbf{r}_i - \mathbf{r}_i^0$, $\mu$ the mobility, and $\mathbf{M}_{ij}$ a $2\!\times\! 2$ block of the dynamical matrix \footnote{We adopt Einstein summation convention}. The orientation $\mathbf{e}_i\!=\!(\cos \theta_i, \sin \theta_i)$ follows the linearized equation 
$\dot \theta_i \!=\! -J (\theta_i -\alpha_i) \!+\! \eta_i$, 
with $\langle \eta_i \rangle\!=\!0$ 
and $\langle \eta_i(t) \eta_j(s)\rangle \!=\! 2 \tau^{-1} \delta_{ij} \delta(t-s)$. Performing the replacement $\theta_i \to \theta_i \!-\! \alpha_i$
and projecting the equations for $\delta \mathbf{r}_i$ onto the normal modes  (see Refs. \cite{henkes2020dense,Bi2016} and the Appendix (\ref{app:EffT}) for details),  we obtain that the mean energy per mode can be written as $e_{\nu} \!=\! \frac{1}{2} T_{eff}^0 I(\tau,J)$ with $T_{eff}^0 \!=\! v_0^2 \tau / 2$. 
For $J\!=\!0$ and $\tau \!\neq\! 0$, it follows a generalization of the equipartition theorem \cite{Maggi14}. Using the expression of the energy per mode, we can 
thus define $T_{eff}(\tau,J)\!=\!T_{eff}^0 I(\tau,J)$.
In general, it is not possible to compute analytically $T_{eff}(\tau,J)$, however, it turns out that it is a decreasing function of $J$ bounded above by $T_{eff}^0$ and below by $T_{eff}^0 / \lambda_\nu \tau$.

We conclude our analysis of the structural properties of 
the tissue by studying the 
features of the liquid state. 
As shown in Fig. (\ref{fig:pd})-(i), where we report the nematic order parameter $S$ (defined in Appendix (\ref{app:obs})), in the liquid phase the system develops weak nematic order. We can thus define two preferred global directions that are individuated by the average direction of the director fields $\mathbf{n}\!=\!\frac{1}{N}\sum_i(\cos2 \alpha_i,\sin 2 \alpha_i)$ computed at a given time step. We indicate with $x_\parallel$ and $x_{\perp}$ respectively
the directions parallel and orthogonal to $\mathbf{n}$. In the nematic phase, the positional order of the cell centers is different along these two directions, as it is shown in Fig. (\ref{fig:pd})-(j) where we report the radial distribution
functions $g_{\parallel}(r)$
and $g_{\perp}(r)$.
The space isotropy is restored in the glassy regime (see panel (l)). The sixfold orientational order replaces the twofold orientational order in the solid phase (see (k) and (h)).

\begin{figure}[!t]
\centering\includegraphics[width=.5\textwidth]{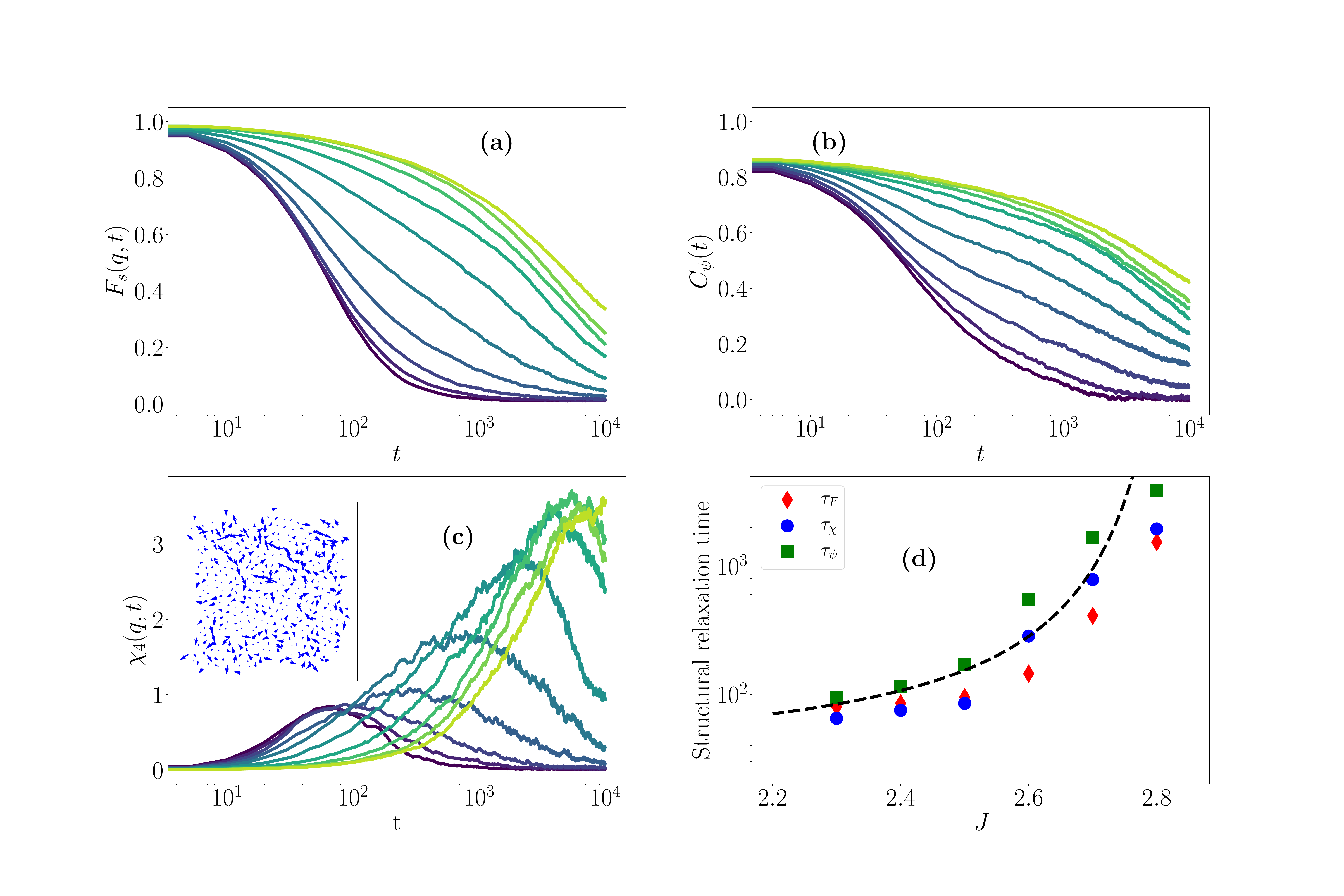}
\caption{ Dynamical slowing down. 
(a) The self-part of the intermediate scattering function
$F(q,t)$ ($J\in[2.3,3.2]$, increasing values from violet to yellow) and corresponding four-point susceptibility $\chi_4(q,t)$ (c). The shape parameters is $s_0\!=\!3.73$ and $T=0.002$.
(b) The correlation function $C_\psi(t)$. 
Inset in (c): Map of displacements for $J\!=\!3.1$ and $t \!\sim\! 10^2$.
(d) Structural relaxation time as a function of the nematic coupling $J$, the dashed black curve is a fit to Vogel-Fulcher-Tammann law with $T \to J^{-1}$. 
}
\label{fig:glassy}      
\end{figure}

\section{Relaxation dynamics}
We now probe the region of the phase diagram where the tissue develops a subdiffusive
regime. Fig. (\ref{fig:glassy})-(a) shows the behavior of
 the intermediate scattering function $F_s(q,t)$ for $s_0\!=\!3.73$, $T=0.002$, and $q\!=\!q_{peak}$ (with $q_{peak}$ the position of the first peak of the static structure factor $S(q)$).
We have also measured the 
time correlation function $C_\psi(t)$ of the hexatic order parameter $\psi_6$ \cite{flenner2015fundamental,massana2018active} (see the Appendix (\ref{app:obs}) for the definition).
The behavior of $C_\psi$ is shown in panel (b) of the same figure. As one can see,
$C_\psi$ undergoes a dynamical slowing down as $J$ increases similar to that observed in $F_s(q,t)$. Since $\psi_6(t)$ reflects the local structure that is determined by the number of cell sides at time $t$, a non-vanishing correlation  $C_\psi(t)$ signals a dynamical slowing down of shape fluctuations.
The dynamical slowing down is usually due to the presence of relaxation dynamics on different time scales.
The emerging of complex and heterogeneous relaxation dynamics becomes more evident 
probing the dynamical susceptibility
$\chi_4(q,t)$
defined as the sample-to-sample fluctuations of 
$F_s(q,t)$, shown in (c) \cite{Berthier11}.
%
$\chi_4(q,t)$ shows a broad peak, due to the presence of dynamical heterogeneity (see the displacement field, inset in (c)), that
grows in height and shifts towards longer times as $J$ increases, the typical feature of glassy systems approaching the glass transition \cite{Berthier11,KobGlotzer1997}. 
We can provide a quantitative measure of the dynamical slowing down using a characteristic relaxation time $\tau_\alpha$ defined as $C(\tau_\alpha)\!=\!e^{-1}$, with $C(t)$ a time-correlation function. In panel (d) we show the behavior of the relaxation time for $F_s(q_{peak},t)$, and
$C_\psi(t)$. Another estimate of $\tau_\alpha$ is provided by the position of the peak of $\chi_4$ (shown in the same panel).
As it has been observed in the case of flocking transition in biological tissues \cite{GiavazziPaoluzzi2018},
the behavior of $\tau_\alpha$ as a function of $J$ proves that the alignment interaction acts as an inverse temperature, causing a cooling down of the system as $J$ increases. This is confirmed by the fact that a Vogel-Fulcher-Tamman formula $\tau_\alpha \propto \exp(B/(J^{-1} - J^{-1}_c))$ (with $T \to J^{-1}$ and $J_c \sim 2.9$) well captures the behavior of the relaxation time.

\section{Discussion and Conclusions}
The collective behavior of biological tissues shows features remarkably similar to those of active nematics, disordered solids, and supercooled liquids. Some of these facts can be rationalized in the framework of dense active matter \cite{Henkes2018}. 
However,
to capture the collective properties of biological tissues in an opportune coarse-grained description one needs to take into account all the relevant ingredients of single-cell dynamics. 
Since moving cells assume an asymmetric configuration that spontaneously breaks spatial symmetries, 
we have studied numerically how feedback between cell shape and displacement changes the structural properties of the tissue. We performed our study
within the framework of the Voronoi models.
We focused our attention on alignment interactions tending to couple the direction of cell motion with its elongation.
Besides the experimental evidence at the single-cell level that highlights the importance of feedback in cell locomotion \cite{trepat2011plithotaxis}, the impact of these interactions on the large-scale behavior of confluent monolayers remains poorly understood. 

We have explored the phase diagram of the tissue using as a control parameter the target shape index $s_0$, which is  experimentally accessible \cite{Park15,Malinverno17}. The second control parameter is the strength of the alignment interaction $J>0$. 
We documented that the interplay of these control 
parameters triggers structural changes 
giving rise to a rich phase diagram characterized 
by liquid-to-solid transitions and glassy dynamics.
For large enough $J$ values, the tissue undergoes 
a phase transition between a disordered state 
and a {\it quasi} hexagonal lattice at  
$s_0 \sim s_0^{hex}$ \cite{staple2010mechanics}. 

For larger $s_0$, the system remains  in a disordered liquid state, showing typical features of glassy dynamics as the strength of the alignment force increases \cite{Angelini11,Bi15,nandi2018random,nandi21}. 
In particular, we observed the proliferation of dynamical heterogeneities, subdiffusive dynamics, broad peaks in the dynamical susceptibilities, and dynamical slowing down of density and shape fluctuations. We showed that the solidification of the system for increasing values of $J$ can be generally understood in terms of an effective temperature $T_{eff}$ that scales with the inverse of $J$, similarly to what has been observed in the case of the self-propelled Voronoi model with polar interactions \cite{GiavazziPaoluzzi2018}.

In conclusion, the intensity of the shape-displacement feedback at the single-cell level can trigger structural transitions in confluent monolayers 
that impact dramatically the collective behavior of the biological tissue.
Our results suggest that, if the coupling between cell elongation and displacement is small, and thus $J$ assumes small values, cells tend to rearrange as in a weakly nematic fluid rather than form a confluent monolayer. In other words, biochemical mechanisms that tend to alter cell polarization might impact the collective properties of the biological tissue.
%
There are biological implications for a weak coupling between shape and locomotion.
For instance, the interplay between these two properties might become 
weak for cells that suddenly lose a polarized shape,
as in the case of metastasis invasion, where the metastatic cell does not show epithelial polarity \cite{paul2017cancer}. Our analysis suggests that, in such a condition, the tissue tends to melt into a fluid phase. More in general and in agreement with recent studies that revealed the crucial role of cell symmetry \cite{jung2019apical}, 
our results show that alignment interactions might provide an additional control parameter for the epithelial-mesenchymal transition.

\subsection*{Acknowledgments}
M.P. has received funding from the European Union's Horizon 2020 research and innovation programme under the MSCA grant agreement No 801370
and by the Secretary of Universities 
and Research of the Government of Catalonia through Beatriu de Pin\'os program Grant No. BP 00088 (2018). I.P. acknowledges MICINN, DURSI and SNSF for financial support under
Projects No. PGC2018-098373-B-I00, No. 2017SGR-884,
and No. 200021-175719, respectively.
M.P. acknowledges Sapienza University of Rome, Physics
Department, for kind hospitality during the preparation of this
manuscript. MCM was supported by the US National Science Foundation though award No. DMR-2041459.


%
%
%
%

\appendix 

\section{Observables} \label{app:obs}
We indicate with $\rr_i(t)$ the position of the $i-$th cell center in the lab reference frame and with $\rr_{i}^\prime(t)$ the position in the center-of-mass reference frame. In the following, we indicate
with $\langle \mathcal{O} \rangle_s$ the average of the observable $\mathcal{O}$ with respect to independent runs, \textit{i.e.}, the subscript $s$ indicates sample averages. 
We indicate with $\langle \mathcal{O} \rangle_t$ time-averaging in the stationary state. 

For studying the single-cell diffusion and the solid-to-liquid transition,
we compute the mean-squared displacement $\Delta r^2$ that is
\begin{equation} \label{eq:msd}
    \Delta r^2 = \frac{1}{N} \left\langle \sum_i \left[ \rr^\prime_i(t) - \rr_i^\prime(0) \right]^2 \right\rangle_s \; . 
\end{equation}
We quantify the emerging of hexatic order
through the complex field $\Psi_i(t)$ defined for each cell
\begin{equation} \label{eq:psit}
\Psi_i(t) = \frac{1}{n} \sum_{j \in n. n.}^n e^{i 6 \theta_{ij}(t) }   
\end{equation}
with $n$ the number of Voronoi neighbors to the cell $i$. The angle $\theta_{ij}$ is individuated by the two cell centers $i$ and $j$.
The hexatic order parameter at the time step $t$ reads 
\begin{equation} \label{eq:psi6}
    \psi_6 (t) = \frac{1}{N} \left| \sum_i \Psi_i(t) \right|
\end{equation}
and we indicate with $\psi_6$ its time average.
We obtain additional and complementary information on
the positional order by measuring the static structure factor
 $\tilde{S}(q_x,q_y)$ that is
\begin{equation} \label{eq:static}
    \tilde{S}(q_x,q_y) = \frac{1}{N} \left\langle \sum_{j,k} e^{i \qv \cdot (\rr_j^\prime - \rr_k^\prime)} \right\rangle_{t,s}
\end{equation}
where the wave vector $\qv=(q_x,q_y)$ satisfies the periodic conditions imposed to the dynamics, \textit{i.e.}, $q_{x,y} = \frac{2 \pi}{L}(n_x,n_y)$, with $n_{x,y} = 0, \pm 1, \pm 2, ...$ (and avoiding the combination $n_x=n_y=0$).

We detect the presence of nematic order measuring the order parameter
\begin{equation} \label{nem_order}
    S = 2 \langle \frac{1}{N} \sum_i \cos^2 \alpha_i \rangle - 1 \; .
\end{equation}
We obtain additional information on the nematic phase measuring
the pair distribution function
\begin{equation} \label{gofr}
    g(\mathbf{r}) = \frac{1}{N} \left\langle \sum_{i,j \neq i} \delta (\mathbf{r} - \mathbf{r}_j + \mathbf{r}_i)\right\rangle_{t,s} \; .
\end{equation}
In particular, we compute $g_{\parallel,\perp}(r)$, where the subscription indicates that
the observable is computed along the principal directions of the nematic director, \textit{i.e.}, $g_\perp(r)\equiv g(0,r_\perp)$ and $g_\parallel(r)\equiv g(r_\parallel,0)$.

As dynamical observables, we measure the self-part of 
the Intermediate Scattering Function $F_s(q,t)$, 
and the time-correlation function of the hexatic order parameter $C_\psi(t)$ \cite{Lacevic2003,Berthier11,flenner2015fundamental,massana2018active}.
The intermediate scattering function is
\begin{equation} \label{eq:fself}
    F_s(q,t) = \frac{1}{N} \left\langle \sum_i e^{ -i \qv \cdot (\rr^\prime_i (t) - \rr^\prime_i(0) ) } \right\rangle_s \; ,
\end{equation}
where $\qv$ follows the same prescription used to compute $S(q_x,q_y)$.
The sample-to-sample fluctuations of $F_s(q,t)$ provides a measure of dynamical
heterogeneity through the four-point dynamical susceptibility $\chi_4(q,t)$.
The position of the peak $\chi_4(q,t)$, \textit{i.e.}, $t=\tau_4$,  individuates the typical time scale of dynamical heterogeneity. We thus compute the displacement field $\Delta \rr (x,y,\tau_4)$\cite{Berthier11}. 
Furthermore, we measure the relaxation time of shape fluctuations using $C_\psi(t)$ defined through the correlation function
\begin{equation} \label{eq:cpsi}
    C_\psi(t) = \frac{1}{C_\psi(0)} \left\langle \sum_i \Psi_i(t) \Psi_i^*(0) \right\rangle_s
    \end{equation}

\section{Phase diagram for $J=0$} \label{app:PDJ0}
\begin{figure*}[!t]
\centering\includegraphics[width=1.\textwidth]{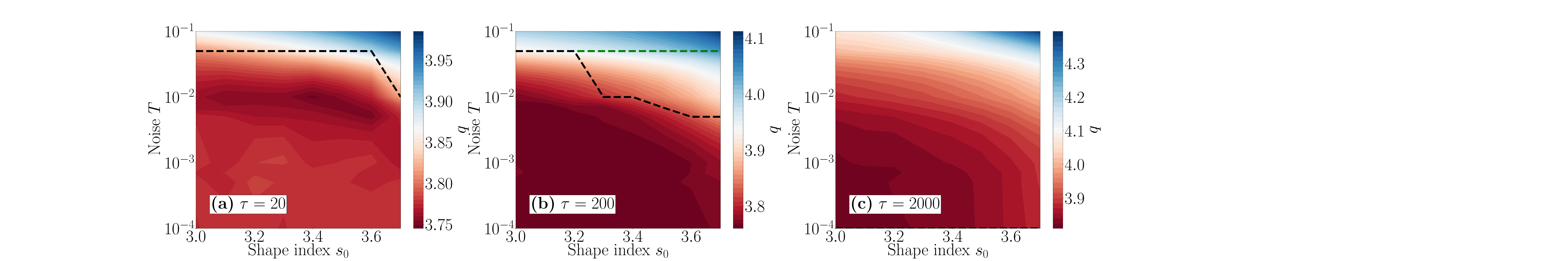}
\caption{ Phase diagram of the model for $J=0$ and $\tau=20,200,2000$, (a), (b), and (c) panel, respectively.
The dashed black line indicates the transition between solid and fluid using the shape parameter $q$ as a structural order parameter. The dashed green line in panel (b) is the region explored in the main text for $J \neq 0$.}
\label{fig:SI1}      
\end{figure*}

In the main text, we have used as control parameters $s_0$ and $J$. For $J=0$, the model reduces to a self-propelled Voronoi model where the self-propulsion is due to correlated noise.  
We have thus probed different regions of the phase diagram for $J=0$ using
as control parameters the strength of the noise $T$ and the shape index $s_0$. The resulting phase diagrams for $\tau=20,200,2000$ and $N=256$ are shown in Fig. (\ref{fig:SI1}). As a structural parameter for discriminating the solid from the fluid phase, we adopt the shape parameter $q$ defined as
\begin{equation}
    q = \left\langle \sum_i \frac{p_i}{\sqrt{a_i}}\;.  \right\rangle
\end{equation}
Following \cite{Bi2016,Bi15,epl}, we use the criterium $q>3.81$ for defining the fluid regime and $q<3.81$ for the solid regime. We explored noise values $T\in[10^{-4},10^{-1}]$ and $s_0\in[3.0,3.7]$. As one can see, for $\tau=200$ the system is always in the fluid state. The region of parameters explored in the main text is highlighted in panel (b). The phase diagram shows the same qualitative features as
for the self-propelled Voronoi model  \cite{Bi2016}.

\section{Collective motion} \label{app:coll}
The presence of migratory patterns has been evaluated using global quantities as the Vicsek order parameter $\Phi$ \cite{Vicsek12} and the nematic order parameter $\Omega$ of the displacements. The order parameter $\Phi$ captures collective cell migration and it is defined as follows
\begin{equation} \label{eq:Phi}
\Phi = \left\langle \frac{1}{N} \left | \sum_k e^{i \phi_k (t)}   \right| \right\rangle_t
    \end{equation}
The order parameter $\Omega$ captures the emerging of nematic order in the cell displacements. It is defined as follows
\begin{equation} \label{eq:Omega}
    \Omega = \left\langle \frac{1}{N} \left | \sum_k e^{i 2 \phi_k (t)} \right| \right \rangle_t
\end{equation}

\begin{figure*}[!t]
\centering\includegraphics[width=1.\textwidth]{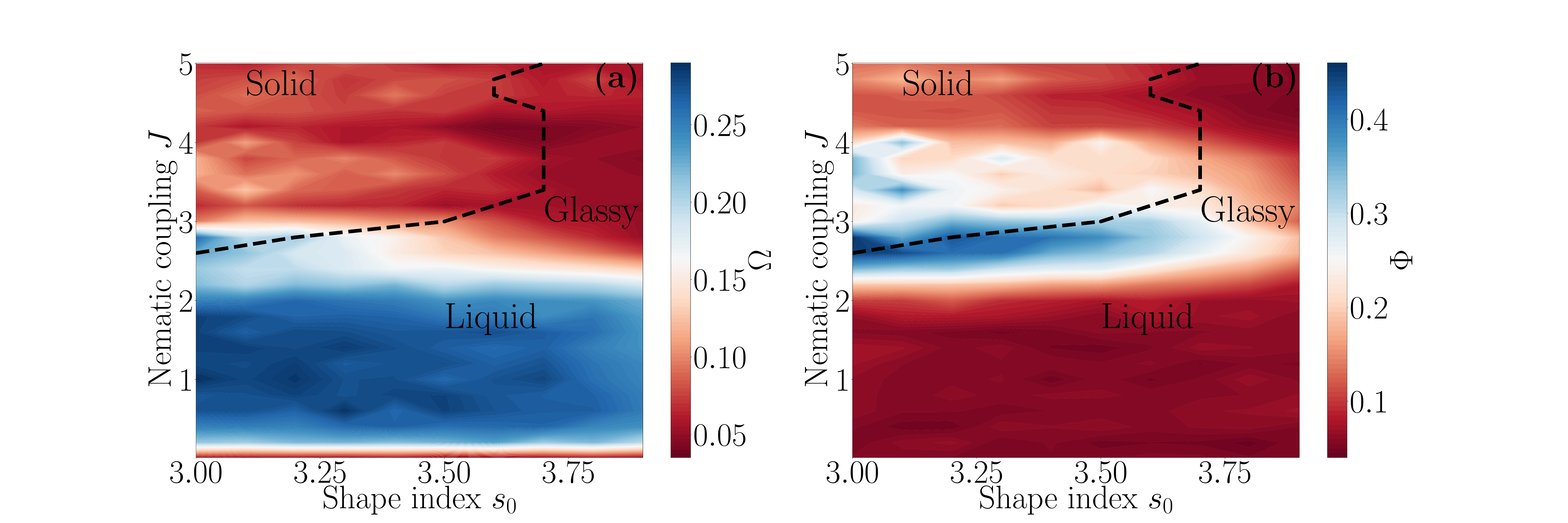}
\caption{ Collective motion. Global nematic order (a) 
and polar order (b) in the velocity field. The dashed black line indicates the transition to solid.}
\label{fig:SI2}      
\end{figure*}

The resulting phase diagrams are shown in Fig. (\ref{fig:SI2}).
The system develops a weak nematic phase in the velocity field in the liquid phase for $0<J<3$, as shown in panel (a). It is worth noting that the order parameter does not overcome the value of $\Omega \sim 0.3$, indicating that only in a small system fraction nematic order is appreciable. In panel (b) we report the behavior of $\Phi$. Around the liquid-solid transition, the system develops weak flocking patterns for $J\sim 3$. In this case, the Vicsek order parameter does not overcome $\Phi \sim 0.4$. 

\begin{figure*}[!t]
\centering\includegraphics[width=1.\textwidth]{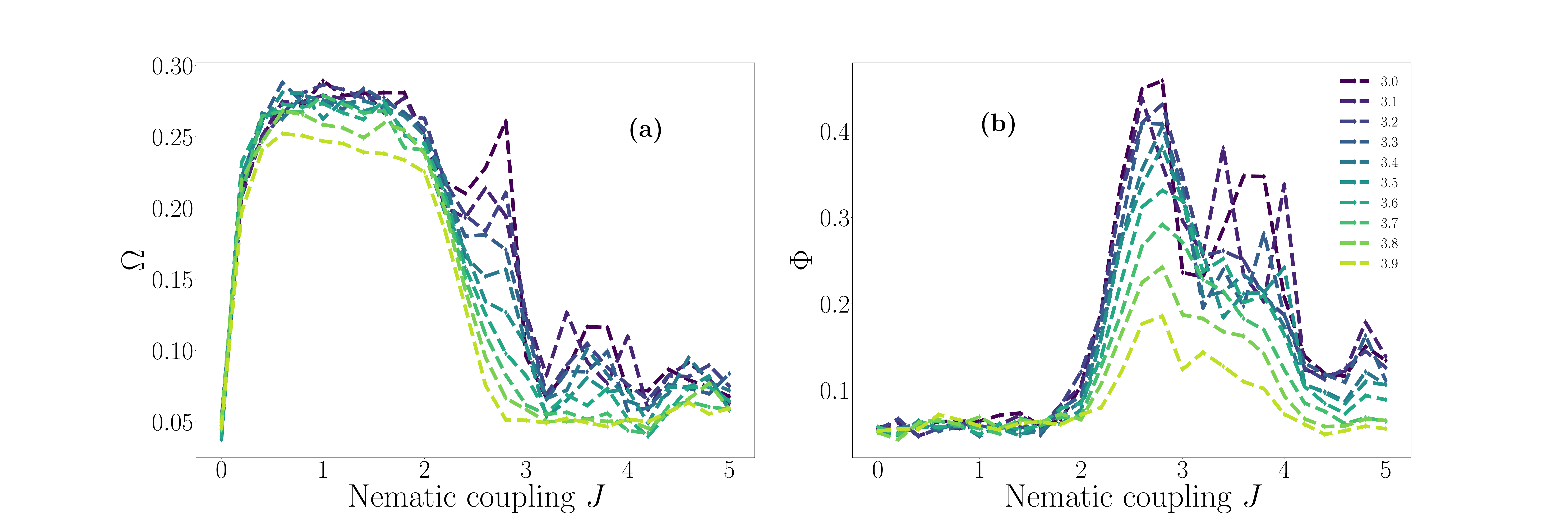}
\caption{ Order parameters $\Omega$ and $\Phi$ as a function of $J$ for increasing values of $s_0$ (from violet to yellow).}
\label{fig:SI3}      
\end{figure*}

The behavior of $\Omega$ and $\Phi$ as a function of $J$ for different values of $s_0$ is shown in Fig.
(\ref{fig:SI3}). The parameter $\Omega$ turns out to be different from zero in the liquid state, almost independently by $s_0$ (panel a). The presence of migratory patterns characterized by polar order ($\Phi \neq 0$), is more evident for small values of $s_0$ (panel b). 

\begin{figure*}[!t]
\centering\includegraphics[width=1.\textwidth]{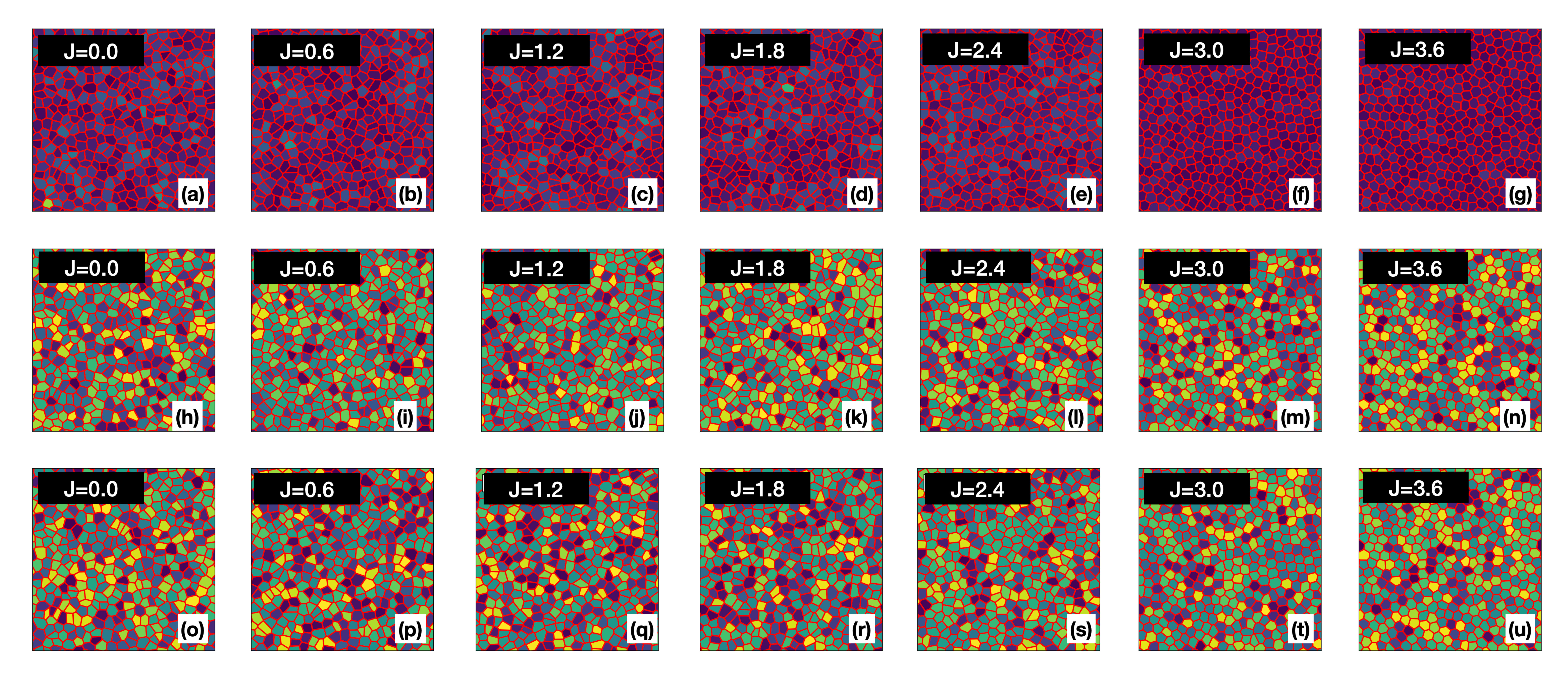} \\
\centering\includegraphics[width=1.\textwidth]{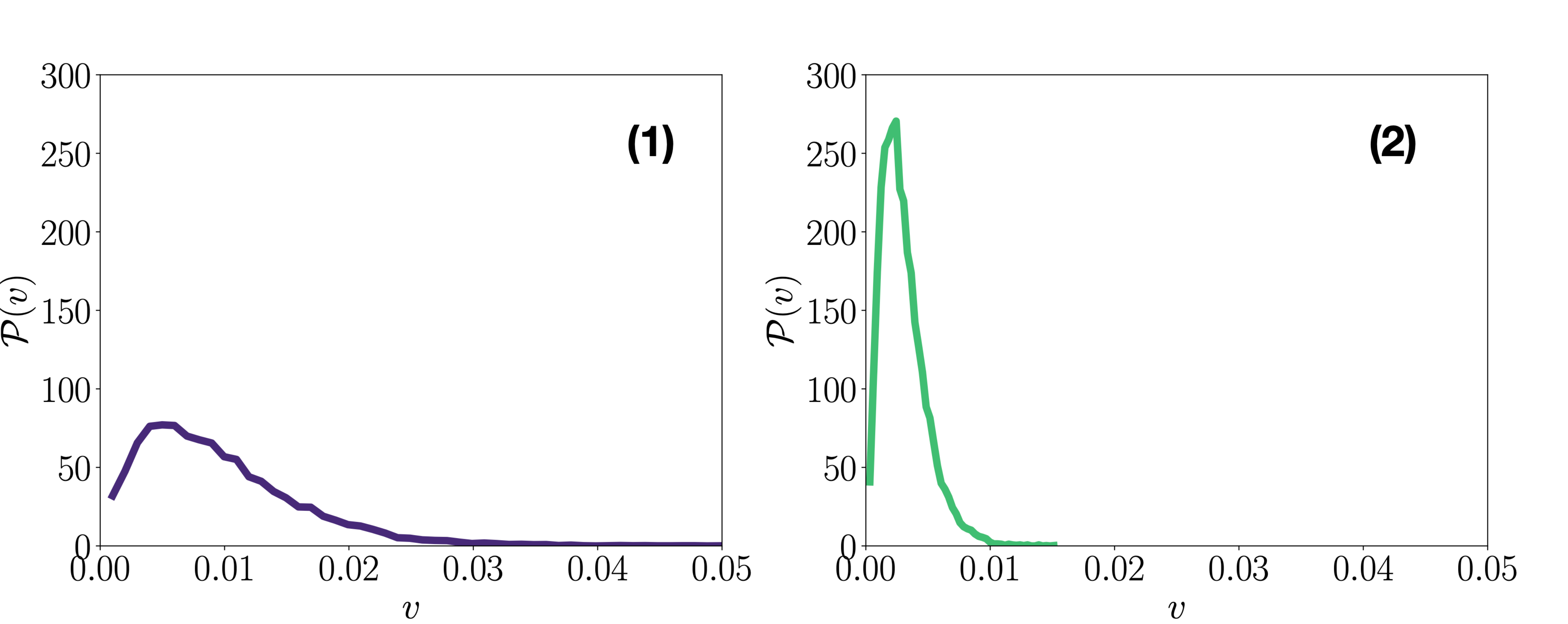}
\caption{ Representative snapshots for $p_0=3.2$ for different values of $J$. In the first row (a-g), the color indicates the modulus of the velocity of each cell. In the second (h-n) and in the third (o-u) row the color indicates nematic and polar angle, respectively, 
obtained from the velocity of the cell and calculated with respect to the $x$-axis. The fourth row (1-2) shows the distribution of the velocity for $J=0.0$ (panel (1)) and $J=3.6$ (panel (2)).}
\label{fig:SI4}      
\end{figure*}

The snapshots of steady-state configurations for $s_0=3.2$ 
are shown in Fig. (\ref{fig:SI4}). In the first row, cells are colored according to their velocity. In the second row, the color indicates the angle of the nematic and polar order, respectively. These parameters have been obtained considering the velocity $\mathbf{v}_i$ of the cell $i$ that can be written as $\mathbf{v}_i=v_i (\cos \theta_i, \sin \theta_i)$. Region of the same color indicates local nematic/polar order. In the fourth row, we report the probability distribution function of the velocity $\mathcal{P}(v)$. $\mathcal{P}(v)$ turns out to be strongly peaked around zero in the solid phase and develops a long tile towards higher values in the liquid state.


\section{Static Structure Factor $S(q_x,q_y)$ and Radial Distribution Function $g(r)$} \label{app:static}
\begin{figure*}[!t]
\centering\includegraphics[width=1.\textwidth]{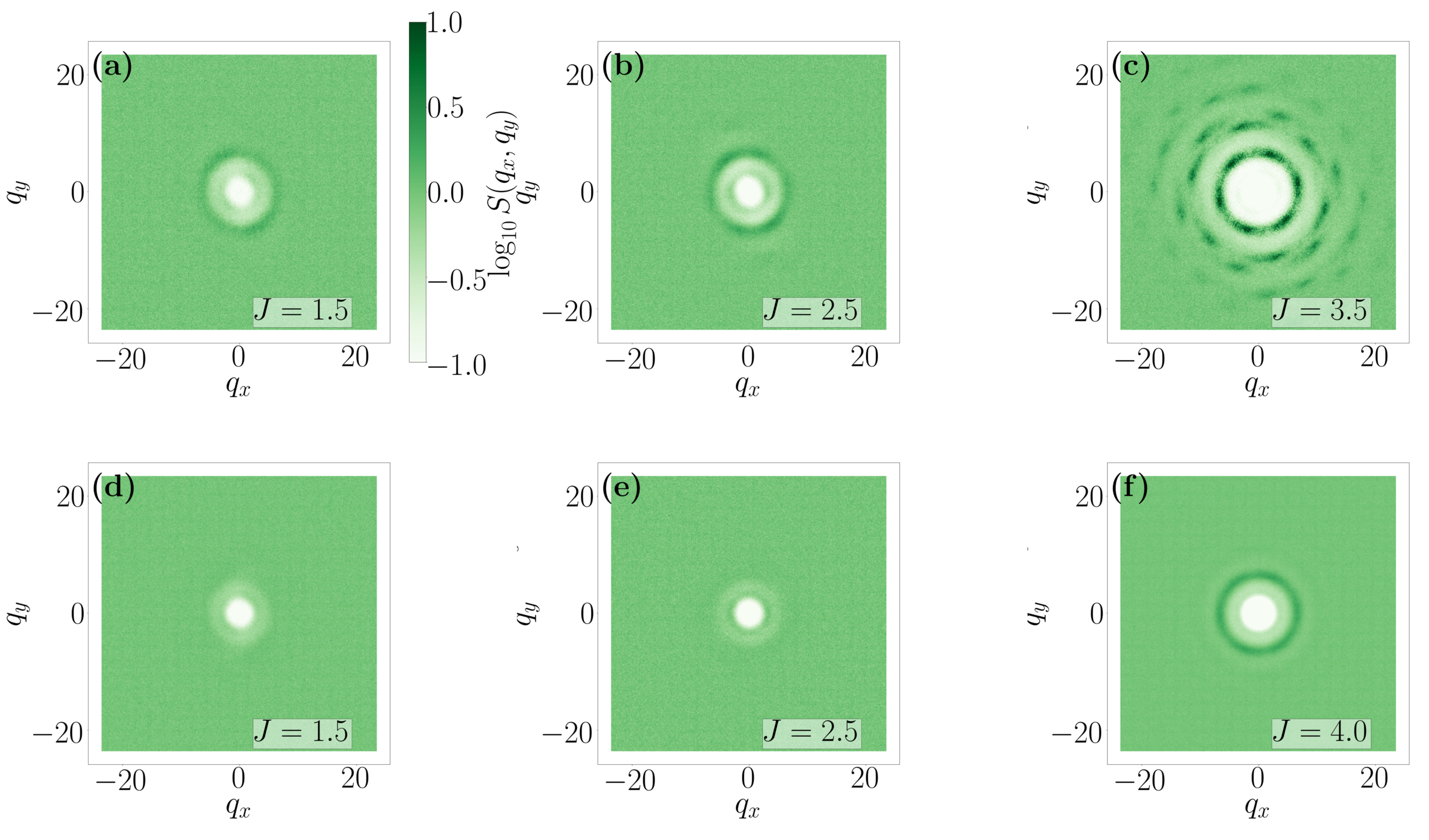}\\
\caption{ Static Structure factor $S(q_x,q_y)$ for $s_0=3.0$ (a,b,c) and $s_0=3.8$ (d,e,f).}
\label{fig:SI5}      
\end{figure*}

In Fig. (\ref{fig:SI5}),
we show the static structure factor $S(q_x,q_y)$. 
Panels (a), (b), and (c) report results 
for $s_0=3.0$ as the intensity
of the alignment interaction grows. 
Crossing the critical value $J \sim 3$, the heat map develops patterns peculiar to the hexatic phase. 
For a larger value of $s_0=3.8$, the system does not undergo a liquid-solid transition anymore.
In this situation, the structure factor does not develop regular peaks (panels (d), (e), and (f)).

\begin{figure*}[!t]
\centering\includegraphics[width=1.\textwidth]{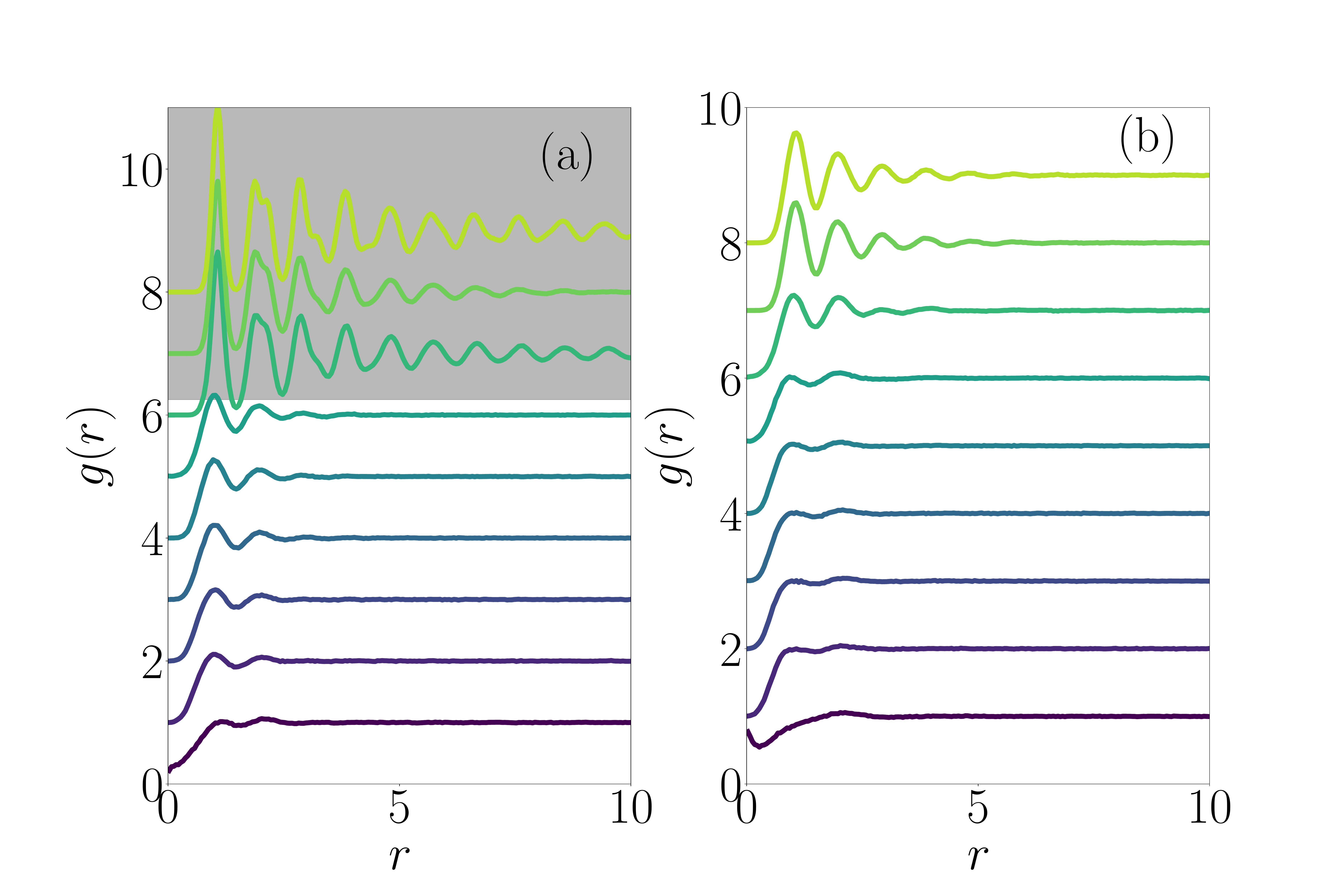}
\caption{ Radial distribution function $g(r)$ for $s_0=3.0$ (a) and $s_0=3.7$ (b). Increasing values of $J=0.0,0.5,1.0,1.5,2.0,2.5,3.0,3.5,5.0$ from violet to yellow, respectively. For clarity curves have been shifted vertically.
The grey area in (a) indicates configurations in the solid phase.}
\label{fig:SI5_gofr}      
\end{figure*}

In Fig. (\ref{fig:SI5_gofr}) we report the radial distribution function $g(r)$ for $s_0=3.0$ and $s_0=3.7$. In both cases, $g(r)$ becomes more structured as the nematic coupling increases. However, while at high $s_0$ values the liquid becomes progressively more structured, at low $s_0$ values when the system crosses from a liquid-like to a solid-like phase, $g(r)$ changes suddenly. The presence of damped oscillations indicates a disordered rather than a crystalline structure.   

\section{Effective temperature with alignment interactions} \label{app:EffT}
We consider a system composed of $N$ self-propelled particles where each of them tends to align towards a given direction defined by the angle $\alpha_i$. The self-propulsion of magnitude $v_0$ acts along $\mathbf{e}_i=(\cos \theta_i, \sin \theta_i)$ and changes direction with a rate $\tau^{-1}$. We indicate with $\mathbf{r}_i^0$ the inherent state configuration that minimizes the mechanical energy of the system. We indicate with $\delta \mathbf{r}_i= \mathbf{r}_i - \mathbf{r}_i^0$ a small displacement around the equilibrium configuration. Linearizing the potential around the minimum of the mechanical energy we obtain the following equations of motion
\begin{eqnarray}
\delta \dot{\mathbf{r}}_i &=& v_0 \mathbf{e}_i + \mu \mathbf{M}_{ij} \delta \mathbf{r}_i \; , \\
\dot{\theta}_i  &=& -J(\theta_i + \alpha_i) + \eta_i \; , \label{eq:angle}
\end{eqnarray}
where the random force satisfies $\langle \eta_i \rangle = 0$ and $\langle \eta_i(t) \eta_j(s) \rangle = 2 \tau^{-1} \delta_{ij} \delta(t-s)$. $\mathbf{M}_{ij}$ is the $2 \times 2 $ block of the dynamical matrix (hereafter we adopt the Einstein summation convention). Expanding the perturbation in terms of the normal modes $\mathbf{u}_i^{\lambda}$ of the dynamical matrix, one has
\begin{equation}
    \delta \mathbf{r}_i = a_\nu (t) \, \mathbf{u}_i^{\nu}
\end{equation}
and the amplitude $a_\nu$ follows the equation of motion
\begin{equation} \label{eq:mode}
 \dot{a}_\nu = -\mu \lambda_\nu a_\nu + \tilde{\eta}_\nu   \; .
\end{equation}
The noise $\tilde{\eta}_\nu$ satisfies
\begin{eqnarray}
\langle \tilde{\eta}_\nu (t) \rangle &=& M(t) \\
\langle \tilde{\eta}_\nu (t) \tilde{\eta}_\mu (s) \rangle &=& \delta_{\mu,\nu} C(t-s) \\
C(t-s) &\equiv& \frac{v_0^2}{2} \langle \cos( \theta(t) - \theta(s)) \rangle \end{eqnarray}
In terms of the probability density function $\rho(\theta,t)$,
the solution of the Fokker-Planck equation generated by (\ref{eq:angle})
with the initial condition $\rho(\theta,t=0)=\delta(\theta - \theta_0)$
reads
\begin{eqnarray}
\rho(\theta,t)&=& \frac{1}{\sqrt{2 \pi \sigma^2(t)}} e^{- \frac{(\theta - M(t))^2}{2 \sigma(t)^2} }    \\ 
\sigma^2(t) &\equiv& \frac{1}{\tau J} (1 - e^{-2 J t}) \\ 
M(t) &\equiv& \theta_0 e^{-J t}
\end{eqnarray}
and thus we can write
\begin{equation}
\langle \cos \Delta \theta \rangle = \cos M(t) e^{-\frac{1}{2} \sigma^2 (t)}
\end{equation}
where we have defined $\Delta \theta \equiv \theta(t) - \theta(0)$ and, without loss of generality we have set $s=0$ and $\theta_0=0$.
Through (\ref{eq:mode}) we can compute the average potential energy stored in each mode
\begin{eqnarray}
e_\nu = \langle \frac{1}{2} \lambda_\nu a_\nu^2\rangle &=& \frac{1}{2} T_{eff} I(\tau,J) \\
T_{eff}^0 &\equiv& \frac{v_0^2 \tau}{2 \mu} \\ 
I(\tau,J) &\equiv& \int_0^\infty \frac{dt}{\tau} e^{-\frac{1}{2} \sigma^2(t) - \mu \lambda_\nu t} \; .
\end{eqnarray}
Where we have introduced the effective temperaure $T_{eff}^0$ that is one of the control parameters of the KMC algorithm. In this way, the equilibrium equipartition theorem is recovered in the limit $\tau=0$. For $\tau \neq 0$ and $J=0$, we recover a generalization of the equipartition theorem, as shown in Ref. \cite{Maggi14} 
\begin{equation}
    e_\nu(J=0) = \frac{1}{2} \frac{T_{eff}^0}{1 + \mu \lambda_\nu \tau} \; .
\end{equation}
Another limiting case is obtained for $J\to \infty$ (and equivalently $\tau \to \infty$), for which
\begin{equation}
    e_\nu(J\to\infty) = \frac{1}{2} \frac{T_{eff}^0}{\lambda_\nu \tau} \; .
\end{equation}
As a consequence, the effective temperature $T_{eff}(\tau, J)$ turns out to be bounded above by $T_{eff}^0$ and decreases towards $T_{eff}^0 / \lambda_\nu \tau $ as $J$ increases.

\section{Cell anisotropy} \label{app:CellAni}
Indicating with $\lambda_{1,2}^i$ the eigenvalues of the shape tensor of the cell $i$ and using the convention $\lambda_1^i > \lambda_2^i$, we define the cell asphericity $\Delta_i$
as
\begin{align}
    \Delta_i = \frac{(\lambda_1^i - \lambda_2^i)^2}{( \lambda_1^i+ \lambda_2^i)^2} \; .
\end{align}
\begin{figure}[!t]
\centering\includegraphics[width=.5\textwidth]{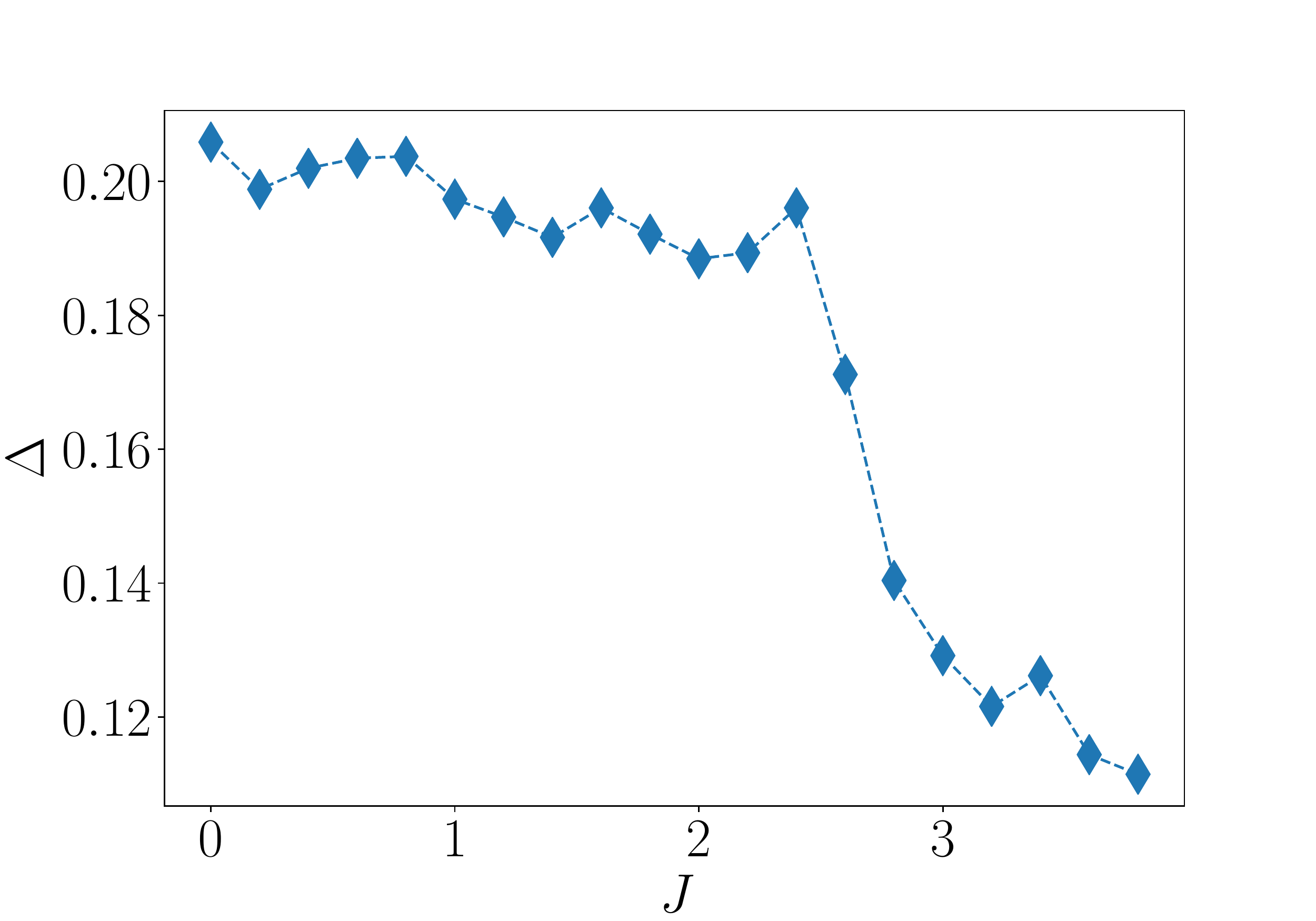}
\caption{ Asphericity $\Delta$ for $s_0=3.0$ as a function of $J$.}
\label{fig:SI_asp1}      
\end{figure}
The asphericity provides a quantitative measure of cell anisotropy, in the case of highly symmetric cells one has $\lambda_1^i \simeq \lambda_2^i$ and thus $\Delta_i \simeq 0$. 
Contrarily, for rod-like cells, $\Delta_i \to 1$, i.e., cells strongly elongated towards a given direction. As shown in Fig. (\ref{fig:SI_asp1}), the parameter $\Delta = \langle \Delta_i \rangle$, where the angular parentheses indicate both averages, over cells and steady-state configurations, 
jumps from higher to lower values crossing the liquid-to-solid transition. We notice that even in the solid phase, where the system arranges in hexagonal patterns, $\Delta \neq 0$, indicating that cells are not displaced in a perfectly regular hexagonal lattice.
\begin{figure}[!t]
\centering\includegraphics[width=.5\textwidth]{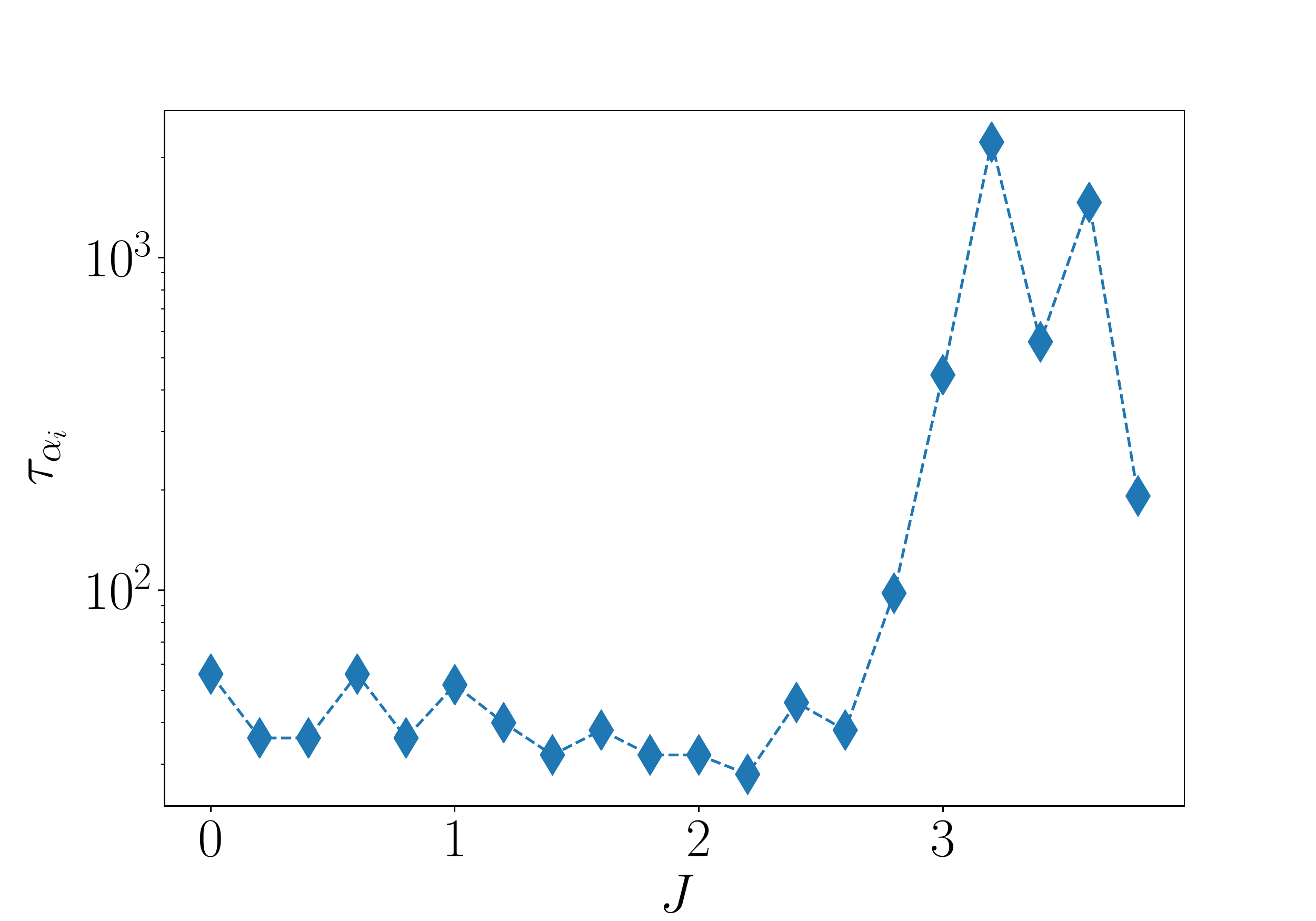}
\caption{ Relaxation time of the direction of maximum elongation for $s_0=1$ as a function of $J$.}
\label{fig:SI_asp_relax}      
\end{figure}
Moreover, shape fluctuations do not flip the principal axes, making the alignment interaction always well defined. For proving this, we measure the relaxation time $\tau_{\alpha_i}$ of the eigenvector corresponding to $\lambda_1^i$. As shown in Fig. (\ref{fig:SI_asp_relax}) the relaxation time is of the order of hundreds of $\tau_{MC}$ in the liquid phase, and it becomes larger in the solid state, indicating that, although the asphericity is small, the direction corresponding to the larger eigenvalue decays on longer times. 
\begin{figure}[!t]
\centering\includegraphics[width=.5\textwidth]{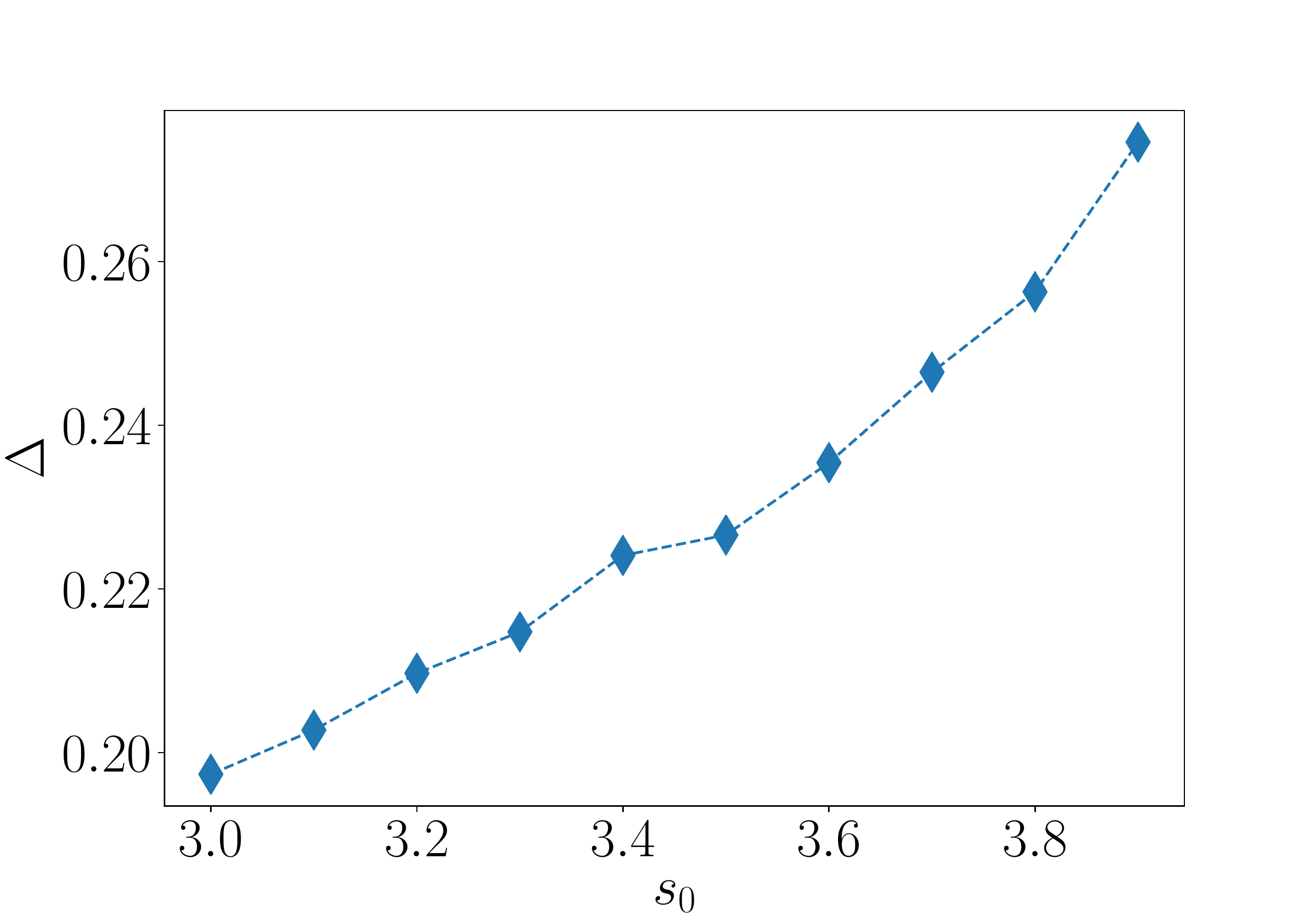}
\caption{ Asphericity $\Delta$ for $J=1$ as a function of $s_0$.}
\label{fig:SI_aspJ}      
\end{figure}
Finally, in Fig. (\ref{fig:SI_aspJ}) we report the behavior of $\Delta$ for $J=1$ as a function of $s_0$. As one can see, although the system does not undergo to structural changes, cells become more elongated making the system more fluid. 

\bibliography{bibflock}
\bibliographystyle{rsc}

\end{document}